\documentclass{article}

\PassOptionsToPackage{numbers, compress}{natbib}

\usepackage[final]{neurips_2020_tda}

\usepackage[utf8]{inputenc} 
\usepackage[T1]{fontenc}    
\usepackage{amsfonts}       
\usepackage{booktabs}       
\usepackage{hyperref}       
\usepackage{url}            
\usepackage{microtype}      
\usepackage{nicefrac}       
\usepackage{paralist}       
\usepackage{siunitx}        
\usepackage{graphicx}
\usepackage{subcaption}
\usepackage{makecell}

\title{
  Characterizing the Latent Space of Molecular Deep Generative Models with Persistent Homology Metrics
}

\author{%
  Yair Schiff$^*$\\
  IBM\\
  Yorktown Heights, NY 10598\\
  \texttt{yair.schiff@ibm.com}\\
  \And
  Vijil Chenthamarakshan\\
  IBM Research\\
  \texttt{ecvijil@us.ibm.com}\\
  \And
  Karthikeyan Natesan Ramamurthy\\
  IBM Research\\
  \texttt{knatesa@us.ibm.com}\\
  \And
  Payel Das\thanks{Corresponding authors}\\
  IBM Research\\
  \texttt{daspa@us.ibm.com}
}

\begin{document}

\maketitle

\begin{abstract}
Deep generative models are increasingly becoming integral parts of the \textit{in silico} molecule design pipeline and have dual goals of learning the chemical and structural features that render candidate molecules viable while also being flexible enough to generate novel designs.
Specifically, Variational Auto Encoders (VAEs) are generative models in which encoder-decoder network pairs are trained to reconstruct training data distributions in such a way that the latent space of the encoder network is smooth.
Therefore, novel candidates can be found by sampling from this latent space.
However, the scope of architectures and hyperparameters is vast and choosing the best combination for \textit{in silico} discovery has important implications for downstream success.
Therefore, it is important to develop a principled methodology for distinguishing how well a given generative model is able to learn salient molecular features.
In this work, we propose a method for measuring how well the latent space of deep generative models is able to encode structural and chemical features of molecular datasets by correlating latent space metrics with metrics from the field of topological data analysis (TDA).
We apply our evaluation methodology to a VAE trained on SMILES strings and show that 3D topology information is consistently encoded throughout the latent space of the model.
\end{abstract}

\section{Introduction}\label{sec:intro}
The discovery process for novel molecules, from ideation to market, typically takes over a decade and costs billions of dollars.
There is therefore a growing need to accelerate this timeline to tackle crises, like COVID-19 or climate change.
Recently, there has been a surge of machine learning (ML), particularly deep learning, methods for learning molecule representations for various applications in chemistry, biology, and materials science.
One specific line of work involves the use of deep generative models to design novel molecules that satisfy certain properties.

These generative models can be trained on a wide variety of inputs, such as image, feature vector, 3D coordinate, text, or graph representations.
The choice of input representation has significant ramifications on performance and success of downstream tasks. 
It also impacts how well different information (e.g. chemical, sub-structural, 3D topological) is captured in a generative model's latent space.
Comparing latent vector representations with a more human-readable and well-established molecular metric can help in estimating the information content of the learned molecular embeddings.

For example, many Natural Language Processing-inspired ML methods for molecular generation train on string representations of molecules known as SMILES \cite{elton2019deep}.
While these 1D string representations have been successfully used in several prediction and design applications, their biggest detractor is that they do not explicitly encode 3D structure, which is important in determining function, such as protein binding affinity or catalyst efficiency.
Finding a way to quantify how much information about 3D structure these generative models can encode, despite having been trained on 1D string representations, will be an important tool in their evaluation and comparison to other models.

In this work, we examine the latent space of a generative model trained on molecular SMILES representations.
We specifically look at the latent space of the VAE model from \citet{chenthamarakshan2020target}, which recently demonstrated success in developing novel, chemically valid, drug-like molecules with potential anti-COVID activity and selectivity.
We compare euclidean distance on latent space vectors from this model and Tanimoto distance on fingerprint representations for the corresponding molecules to see how well molecular sub-structure information is captured.
The novelty of our work comes from our evaluation of how well the VAE's latent space encodes 3D topological structure of molecules.
We perform this evaluation by correlating latent space euclidean distance (referred to throughout as $z$ distance) with a metric on two parameter persistent homology of molecules.
We explore two parameter persistent homology both because of its recent success in virtual molecular screening \cite{keller2018persistent} and because of its ability to simultaneously capture structural and chemical features, by way of the second parameter.
We find that for the VAE from \citet{chenthamarakshan2020target}, 3D information is captured in a uniform manner across the latent space.
In contrast, information from the more common bit vector fingerprint representation, is well captured in regions of the latent space that are near the training distribution, but not consistently across the entire latent space.
We believe that this work provides a means of connecting generative models trained on inputs that do not explicitly encode 3D information with representations of 3D molecular topology.

\section{Methodology}\label{sec:methodology}
The basic tenet of our methodology is to quantify how well the latent space of a generative model encodes semantic information about the underlying data.
We achieve this by correlating the $z$ distance with metrics on other molecular representations, such as persistence diagrams and fingerprints.

The persistent homology of a distance function of a molecule can be calculated by treating the constituent atoms' 3D coordinates as a discrete point cloud.
A second parameter (e.g. atomic partial charge) can be used to control the number of atoms in the point cloud, resulting in two parameter persistent homology.
We compute the Restricted Hilbert function \cite{keller2018persistent}, which evaluates the Betti numbers at each step of the bifiltration.
We use the $\ell_2$ distance between the Restricted Hilbert functions of two molecules as a distance metric between their two parameter persistence diagrams.
For more details, please see Appendix \ref{app:two_param}, and for more technical details about the relevant terminology and calculations, please see \citet{keller2018persistent}.
Our pipeline for calculating two parameter persistent homology on molecules is as follows: first, the RDKit \cite{landrum2013rdkitsoftware, landrum2013rdkit} and OpenBabel \cite{o2011opensoftware, o2011open} libraries are used to convert SMILES representations to 3D atomic point clouds and to extract parameters, such as partial charge and atomic radius.
For each molecule, its two parameter persistence diagram and Restricted Hilbert function are calculated with the RIVET tool \cite{lesnick2015interactive, rivet}.

To have a benchmark for how well the latent space encodes information for other representations, we compare it with the Tanimoto distance on fingerprint representations of molecules.
Fingerprints are widely used bit vector representations that encode information about sub-structures present in a molecule \cite{elton2019deep}.
We use the RDKit library to generate molecular fingerprints and to calculate Tanimoto distance, which measures alignment between the bit vectors of two molecules (see Appendix \ref{app:fp} for more information).

While our approach for evaluating the latent space of generative models is not tied to any specific architecture or implementation, for the purposes of this work we use the VAE model and training detailed in \citet{chenthamarakshan2020target}.
One of the strongest indications that the latent space of this model is able to encode 3D structure to some extent is the success of the model in generating molecules capable of binding to the 3D pocket of target protein structures \cite{chenthamarakshan2020target}.
We use both the latent embeddings of a subset of the training data, as well as randomly sampled vectors from the latent space distribution, which are in turn decoded back to SMILES strings.
For the training data embeddings, we simply use their corresponding SMILES for the persistent homology pipeline described above.
For the random vectors drawn from the latent distribution, we pass them through the VAE decoder network to obtain corresponding SMILES strings and use the RDKit library to discard non-meaningful samples.
The purpose of also analyzing random vectors drawn from the latent distributions is to see how well the latent space encodes information in regions that are not near training data.

To capture an overview of the latent space of this model as a whole, we follow an approach similar to that described in \citet{maragakis2020deep}.
First, we calculate all pairwise $z$ distances for $N$ random data points.
This produces $\mathcal{O}(N^2)$ pairwise distances.
These distances are quantized into $B$ bins of equal width.
We then sample, $n$ pairwise distances from each of the bins, which gives us $B*n$ random latent embedding pairwise distances.
For each of the molecules pairs that correspond to the $B*n$ distances, we also calculate the $\ell_2$ Restricted Hilbert function and Tanimoto distances.
Plotting these metrics and calculating their correlation coefficient with the $z$ distances provides an overview of how well the latent space is able to encode different types of structural information from the molecules.

\section{Related work}\label{sec:related_work}
Correlating metrics on outputs and embeddings of deep generative models with metrics on other molecule representations is a widely used technique and is a natural approach for quantifying and providing a graphical representation of how well a model captures semantic information.
However, to the best of our knowledge, in the domain of molecular generative models, most works rely on metrics on the 1D fingerprint representations of molecules, such as Jaccard or Tanimoto distance.

For example, in \citet{duvenaud2015convolutional}, where the authors create a differentiable fingerprint representation for molecules, they employ this approach by seeing how Tanimoto distance on canonical circular fingerprint representations of molecules correlates with this same metric on their neural network generated fingerprint representation.
Similarly, a graph like the ones presented in Appendix \ref{app:supp} comes from \citet{maragakis2020deep} where the authors compare latent space euclidean distance with Jaccard distance on fingerprint representations.
Our approach is novel in that we compare latent space distances to a more informative and useful measure, namely distances between two parameter persistence diagrams.

An explicit use of persistent homology to evaluate generative models in other domains, such as tabular data, is presented in \citet{charlier2019phom}.
However, the analysis in that work is quite different from ours, since \citet{charlier2019phom} compare persistent homology of data distributions, training vs. generated, whereas we calculate the persistent homology of 3D atomic coordinates of individual molecules, following \citet{keller2018persistent}.

\section{Results}\label{sec:results}
As mentioned in Section \ref{sec:methodology}, in our analysis we use both samples from the training data and randomly sampled points from the latent space.
For the training data, we start with 10,000 training molecules.
We calculate the $\sim$50 million pairwise $z$ distances for these 10,000 points and bin these distances into 10 equal-width groups.
From each of these bins, we randomly select 400 molecule pairs.
The 10th bin is excluded, since it does contain at least 400 points, giving us 3,600 pairwise distances.
For each pair of molecules, we then calculate the $\ell_2$ distance on Restricted Hilbert functions and Tanimoto distance.
Restricted Hilbert functions are calculated for two separate bifiltrations, one where atomic partial charge is used as the second parameter, and one where atomic radius is the second parameter.
For these 3,600 molecule pairs, we calculate the \textit{Pearson r} correlation coefficient between distances.

We repeat this process for random vectors sampled from the VAE latent distribution to see how well information is encoded in the latent space away from the training data.
Specifically, we sample 5,000 vectors from the VAE latent distribution.
Of these, about 4,500 have valid SMILES string decodings.
As above, we calculate all the pairwise $z$ distances for the 4,500 latent space vectors and create ten equal-width bins for these $\sim$10 million distances.
From each distance bin, we randomly choose up to 400 pairs and repeat the correlation calculations between latent and fingerprint / topological representation metrics.
Again, the 10th bin is excluded, since it does contain at least 400 points, giving us 3,600 pairwise distances.

\begin{figure}
    \begin{subfigure}{0.49\textwidth}
        \includegraphics[width=\linewidth]{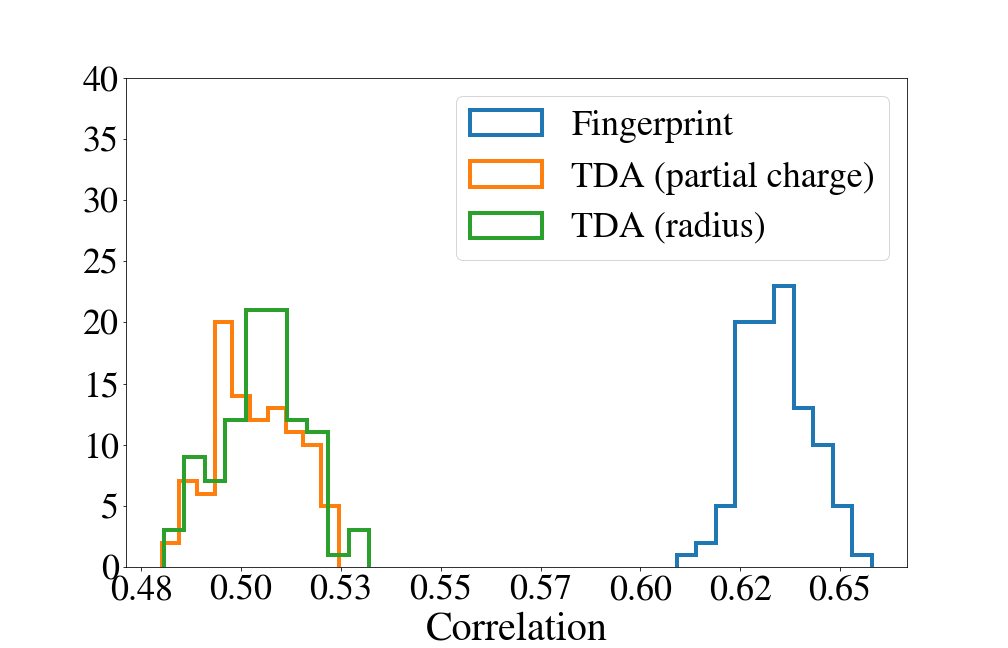}
        \caption{Training data points}
    \end{subfigure}
    \hspace{0.5em}
    \begin{subfigure}{0.49\textwidth}
        \includegraphics[width=\linewidth]{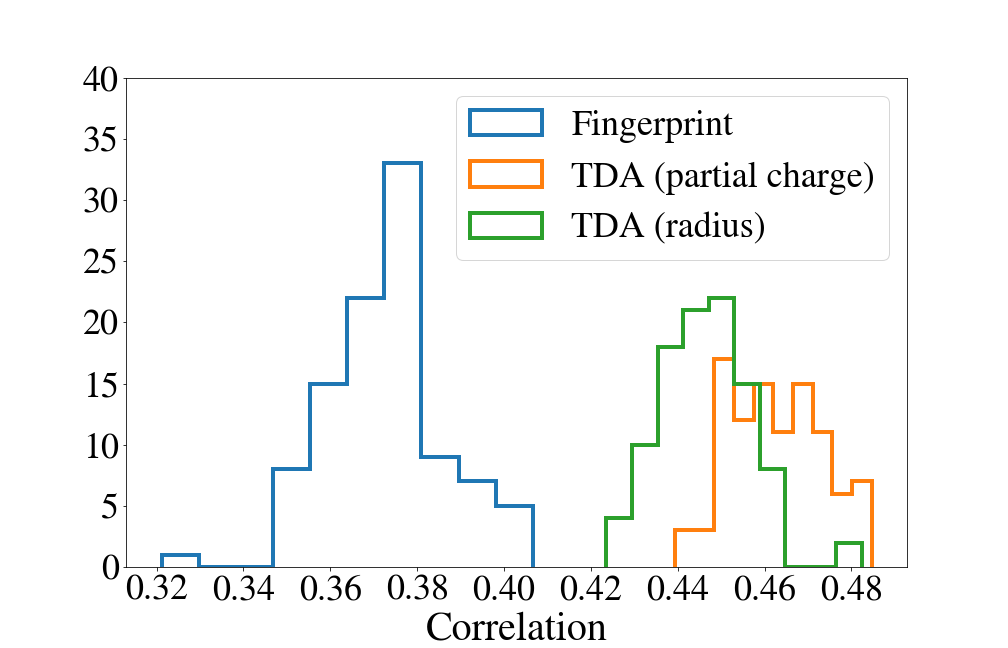}
        \caption{Random samples latent distribution}
    \end{subfigure}
    \caption{Distributions of correlation coefficients
    for 100 random draws of 400 pairs from each distance bin for (a) training molecules and (b) synthesized SMILES strings corresponding to molecules sampled randomly using the latent space distribution.
    For each of the 100 random draws, correlations are calculated between the 3,600 pairwise $z$ distances and Tanimoto distances, $\ell_2$ Restricted Hilbert function (TDA) distances when partial charge is used as the second parameter, and $\ell_2$ Restricted Hilbert function (TDA) distances when atomic radius is used as the second parameter.}
    \label{fig:corr_dist}
\end{figure}

To get robust estimates of this correlation, we repeat the process of randomly drawing 400 pairs from each distance bin 100 times.
For each of these draws, we calculate the correlation coefficient between $z$ distances and metrics for the other molecule representations.
The distributions of these correlations are presented in Figure \ref{fig:corr_dist} and their corresponding statistics are recorded in Table \ref{tab:corr_dist}.
\begin{table}
  \caption{Statistics for distributions in Figure \ref{fig:corr_dist}.}
  \label{tab:corr_dist}
  \centering
  \setlength\extrarowheight{-12pt}
  \begin{tabular}{lccc|ccc}
    \toprule
    \multicolumn{1}{c}{} & \multicolumn{3}{c}{\textit{Training data}} & \multicolumn{3}{c}{\textit{Random latent sample}}\\
    \cmidrule(r){2-4} \cmidrule(r){5-7}\\
    \multicolumn{1}{c}{} & \thead{Fingerprint} & \thead{TDA\\(partial charge)} & \thead{TDA\\(radius)} & \thead{Fingerprint} & \thead{TDA\\(partial charge)} & \thead{TDA\\(radius)}\\
    \midrule
    Median    & 0.635 & 0.503 & 0.507 & 0.377 & 0.464 & 0.449\\
    Mean      & 0.636 & 0.504 & 0.506 & 0.377 & 0.465 & 0.449\\
    Std. dev. & 0.008 & 0.010 & 0.010 & 0.014 & 0.010 & 0.011\\
    \bottomrule
  \end{tabular}
\end{table}

Figure \ref{fig:corr_dist}(a) reveals that for the training data points, $z$ distance shows stronger correlation with Tanimoto distance, when compared to distance estimated using two parameter persistent homology representation.
This result is not surprising, since the VAE model is trained to minimize reconstruction loss for the training distribution.
For training samples, sub-structure information that is explicitly encoded by both the SMILES string and fingerprint representations is better preserved in the latent space.
However, while investigating the random samples generated from the latent space, we find that the correlation between $z$ distance and Tanimoto distance is significantly weaker. 
In contrast, the correlation between $z$ distance and  persistent homology-based distance is consistently around $0.5$ for both training molecules and random samples generated from the latent distribution.
This result indicates that the 3D topological information is moderately, but uniformly, captured throughout the latent space, when compared to Tanimoto distance.
Figure \ref{fig:corr_dist} further shows that the correlations with $\ell_2$ Restricted Hilbert function distance are consistent for both of the second parameters we examine, partial charge and atomic radius, indicating that the VAE latent space is able to consistently capture the 3D topology of molecules, irrespective of the choice of the second parameter.
Figures \ref{fig:2d_diagram}, \ref{fig:3d_diagram}, and \ref{fig:hom_diagram} in Appendix \ref{app:supp} show one example of a molecule pair for which the $z$ distance correlates better with the persistent homology-based distance, when compared to Tanimoto distance. 
\section{Conclusion and future directions}\label{sec:future_work}
TDA tools are being adopted across ML and other domains of science, and we believe this is an exciting direction for \textit{in silico} molecular screening with deep generative models.
Our approach is useful for comparing the semantics of latent spaces of different architectures and, although not presented here, for tracking how the latent space of a particular model evolves throughout training in its ability to learn 3D molecular structures.
One immediate extension of our work includes developing new visualizations that compare latent metrics to metrics on topological structures.
We will also extend this analysis to investigate the effect of using different fingerprints and different generative models on the final results.
More broadly, we believe that incorporating topological metrics directly into the generative model training pipeline can significantly improve the models.
However, this will be a non-trivial effort requiring significant research due to the computationally intensive and non-differentiable pipeline for computing two parameter persistent homology.


\bibliographystyle{abbrvnat}
\bibliography{refs.bib}

\begin{thebibliography}{15}
\providecommand{\natexlab}[1]{#1}
\providecommand{\url}[1]{\texttt{#1}}
\expandafter\ifx\csname urlstyle\endcsname\relax
  \providecommand{\doi}[1]{doi: #1}\else
  \providecommand{\doi}{doi: \begingroup \urlstyle{rm}\Url}\fi

\bibitem[o20()]{o2011opensoftware}
The open babel package.
\newblock URL \url{http://www.openbabel.org}.

\bibitem[Charlier et~al.(2019)Charlier, State, and Hilger]{charlier2019phom}
J.~Charlier, R.~State, and J.~Hilger.
\newblock Phom-gem: Persistent homology for generative models.
\newblock In \emph{2019 6th Swiss Conference on Data Science (SDS)}, pages
  87--92. IEEE, 2019.

\bibitem[Chenthamarakshan et~al.(2020)Chenthamarakshan, Das, Padhi, Strobelt,
  Lim, Hoover, Hoffman, and Mojsilovic]{chenthamarakshan2020target}
V.~Chenthamarakshan, P.~Das, I.~Padhi, H.~Strobelt, K.~W. Lim, B.~Hoover, S.~C.
  Hoffman, and A.~Mojsilovic.
\newblock Target-specific and selective drug design for covid-19 using deep
  generative models.
\newblock \emph{arXiv preprint arXiv:2004.01215}, 2020.

\bibitem[Cignoni et~al.(2008)Cignoni, Callieri, Corsini, Dellepiane, Ganovelli,
  and Ranzuglia]{LocalChapterEvents:ItalChap:ItalianChapConf2008:129-136}
P.~Cignoni, M.~Callieri, M.~Corsini, M.~Dellepiane, F.~Ganovelli, and
  G.~Ranzuglia.
\newblock {MeshLab: an Open-Source Mesh Processing Tool}.
\newblock In V.~Scarano, R.~D. Chiara, and U.~Erra, editors, \emph{Eurographics
  Italian Chapter Conference}. The Eurographics Association, 2008.
\newblock ISBN 978-3-905673-68-5.
\newblock
  \doi{10.2312/LocalChapterEvents/ItalChap/ItalianChapConf2008/129-136}.

\bibitem[Duvenaud et~al.(2015)Duvenaud, Maclaurin, Iparraguirre, Bombarell,
  Hirzel, Aspuru-Guzik, and Adams]{duvenaud2015convolutional}
D.~K. Duvenaud, D.~Maclaurin, J.~Iparraguirre, R.~Bombarell, T.~Hirzel,
  A.~Aspuru-Guzik, and R.~P. Adams.
\newblock Convolutional networks on graphs for learning molecular fingerprints.
\newblock In \emph{Advances in neural information processing systems}, pages
  2224--2232, 2015.

\bibitem[Elton et~al.(2019)Elton, Boukouvalas, Fuge, and Chung]{elton2019deep}
D.~C. Elton, Z.~Boukouvalas, M.~D. Fuge, and P.~W. Chung.
\newblock Deep learning for molecular design—a review of the state of the
  art.
\newblock \emph{Molecular Systems Design \& Engineering}, 4\penalty0
  (4):\penalty0 828--849, 2019.

\bibitem[Keller et~al.(2018)Keller, Lesnick, and Willke]{keller2018persistent}
B.~Keller, M.~Lesnick, and T.~L. Willke.
\newblock Persistent homology for virtual screening.
\newblock \emph{ChemRxiv}, 2018.

\bibitem[Landrum()]{landrum2013rdkitsoftware}
G.~Landrum.
\newblock Rdkit: Open-source cheminformatics.
\newblock URL \url{https://www.rdkit.org}.

\bibitem[Landrum(2013)]{landrum2013rdkit}
G.~Landrum.
\newblock Rdkit: A software suite for cheminformatics, computational chemistry,
  and predictive modeling, 2013.

\bibitem[Lesnick and Wright(2015)]{lesnick2015interactive}
M.~Lesnick and M.~Wright.
\newblock Interactive visualization of 2-d persistence modules.
\newblock \emph{arXiv preprint arXiv:1512.00180}, 2015.

\bibitem[Maragakis et~al.(2020)Maragakis, Nisonoff, Cole, and
  Shaw]{maragakis2020deep}
P.~Maragakis, H.~Nisonoff, B.~Cole, and D.~E. Shaw.
\newblock A deep-learning view of chemical space designed to facilitate drug
  discovery.
\newblock \emph{arXiv preprint arXiv:2002.02948}, 2020.

\bibitem[O'Boyle et~al.(2011)O'Boyle, Banck, James, Morley, Vandermeersch, and
  Hutchison]{o2011open}
N.~M. O'Boyle, M.~Banck, C.~A. James, C.~Morley, T.~Vandermeersch, and G.~R.
  Hutchison.
\newblock Open babel: An open chemical toolbox.
\newblock \emph{Journal of cheminformatics}, 3\penalty0 (1):\penalty0 33, 2011.

\bibitem[{Schr\"odinger, LLC}(2015)]{PyMOL}
{Schr\"odinger, LLC}.
\newblock The {PyMOL} molecular graphics system, version~1.8.
\newblock November 2015.

\bibitem[{The RIVET Developers}()]{rivet}
{The RIVET Developers}.
\newblock Rivet.
\newblock URL \url{https://github.com/rivetTDA/rivet}.

\bibitem[Zhou et~al.(2018)Zhou, Park, and Koltun]{Zhou2018}
Q.-Y. Zhou, J.~Park, and V.~Koltun.
\newblock {Open3D}: {A} modern library for {3D} data processing.
\newblock \emph{arXiv:1801.09847}, 2018.

\end{thebibliography}

\newpage
\appendix
\section{Two parameter persistence homology}\label{app:two_param}
We briefly describe persistent homology in the context of molecular data.
A more detailed presentation can be found in \citet{keller2018persistent}.
First, the 3D atomic coordinates of a molecule are treated as a discrete point cloud.
Using a filtration, such as Vietoris-Rips, sequences of simplicial complexes are created by successively connecting atoms that are further away from each other.
For each step in the filtration, the homology of dimensions zero, one, and two can be thought of as representing the connected components, holes, and voids (respectively) of the simplicial complex, and the the rank or Betti numbers $\mathcal{B}_i$ for $i \in 0, 1, 2$, can be thought of as a count of these different dimensional holes.
In two parameter persistent homology, the second parameter acts as a threshold for which points are considered in the point cloud.
That is, as the threshold for the second parameter varies, more and more atoms are included in the simplicial complex filtration.
The Restricted Hilbert function evaluates $\mathcal{B}_i$ at each step of the bifiltration.
$\ell_2$ distance for two functions is defined as:
$$\ell_2(f, g) = \sqrt{\int(f-g)^2dA}$$
In this work, we evaluate only 0th and 1st dimension homologies both for computational considerations and because, as noted in \citet{keller2018persistent}, the majority of the energy for $\ell_2$ distances on Restricted Hilbert functions for molecules comes from lower dimension homologies.

\section{Fingerprint representations}\label{app:fp}
The fingerprint representations and related metrics used in this work are evaluated using the RDKit library.
First SMILES strings, which use ASCII characters to encode molecules' graph structure \cite{elton2019deep}, are converted to bit vectors know as Molecular ACCess System (MACCS) keys fingerprints.
In these MACCS fingerprint signatures, each bit records a binary structural feature of the molecule \cite{elton2019deep}.

Tanimoto similarity measures the alignment between two bit vector signatures by taking a ratio commonly known as `intersection over union' in other contexts.
Specifically, it is calculated as the bit-wise \texttt{AND} ($\cap$) divided by bit-wise \texttt{OR} ($\cup$) of two bit vectors:
$$Tanimoto(A,B) = \frac{A \cap B}{A \cup B}$$
Subtracting this quantity from 1 gives a distance rather than a similarity metric.

\section{VAE Model}\label{app:vae}
In this section, we provide a brief overview of the VAE model that we examine, which comes from \citet{chenthamarakshan2020target}.
In \citet{chenthamarakshan2020target}, the authors begin with semi-supervised training of the encoder-decoder VAE networks on $\sim$2 million molecules.
VAE loss is comprised of an L2 reconstruction term and KL divergence annealing to ensure smoothness of the latent space.
For both the encoder and decoder, the authors use Gated Recurrent Unit architectures.
Latent space vectors are 128-dimensional.
In addition to optimizing reconstruction loss and KL loss, the latent embeddings are jointly trained on two property prediction tasks.
To generate molecule candidates, the authors use an efficient sampling scheme to search for molecules that satisfy several, often conflicting, specifications.
For more information about the architecture, hyperparameter selection, training protocol, and results, please see \citet{chenthamarakshan2020target}.

We also provide more detail here about how training samples and randomly sampled latent vectors are used in our evaluation pipeline.
Because our analysis compares metrics for multiple molecule representations, for each sample, we needed its latent vector, bit vector fingerprint, and two parameter persistence diagram representations.

For the training samples, for each data point, we start with the SMILES string.
This string is passed through the VAE's encoder network to obtain the corresponding latent vector.
The SMILES string is also processed by the the RDKit library to produce its MACCS fingerprint signature.
Finally, the OpenBabel library is used to extract 3D coordinates of the molecule's constituent atoms, which are saved as a point cloud and passed through the RIVET tool to produce persistence diagrams.

For the randomly sampled latent vectors, we begin by first drawing high-dimensional vectors from the encoder's learned latent distribution.
These vectors are passed through the VAE's decoder network to produce SMILES strings from which we can produce corresponding fingerprint and persistence diagram representations.

\section{Supplementary correlation plots}\label{app:supp}
\begin{figure}[!htb]
    \centering
    \includegraphics[width=\textwidth]{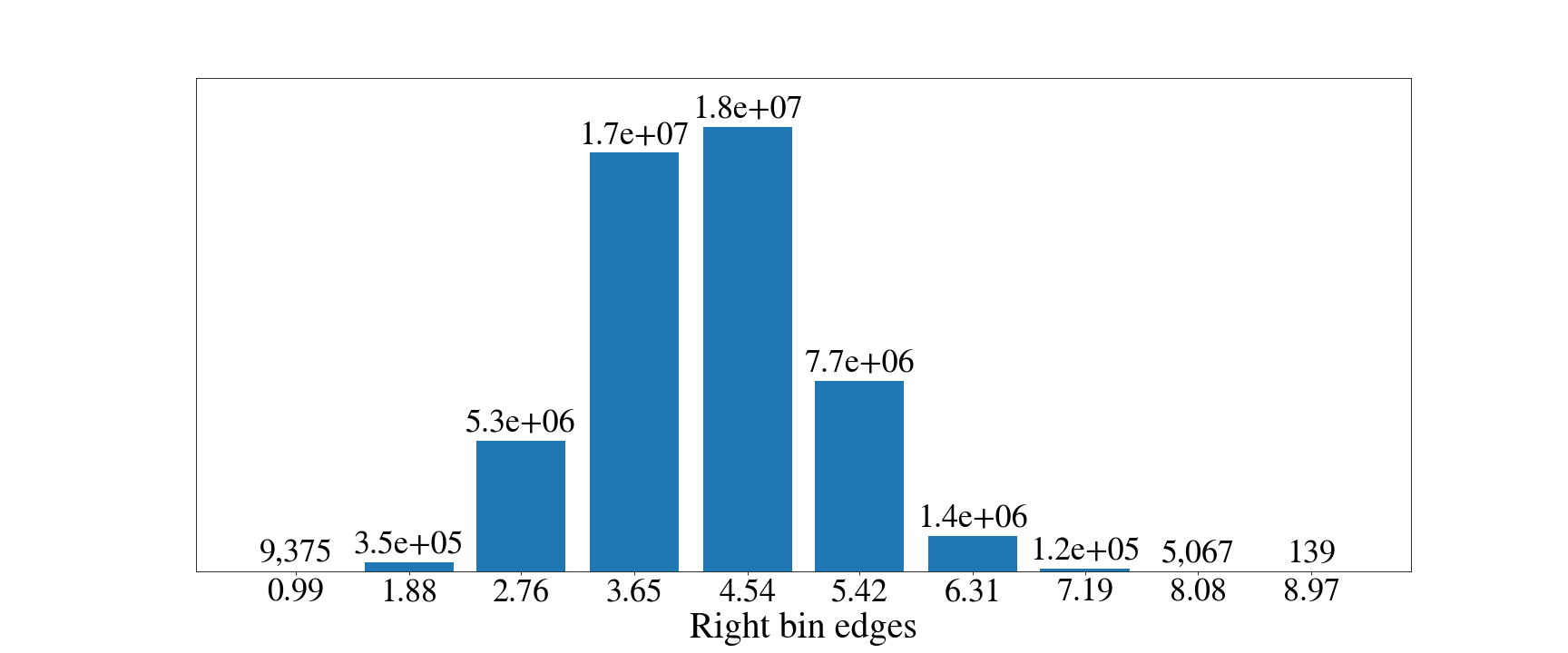}
    \caption{Number of pairwise $z$ distances within each bin for 10,000 training molecules.
    Horizontal axis marks represent the right edge of each bin.
    Data labels represent the count of pairwise distances that fall within each bin.}
    \label{fig:bin_dist_training}
\end{figure}

\begin{figure}[!htb]
    \centering
    \includegraphics[width=\textwidth]{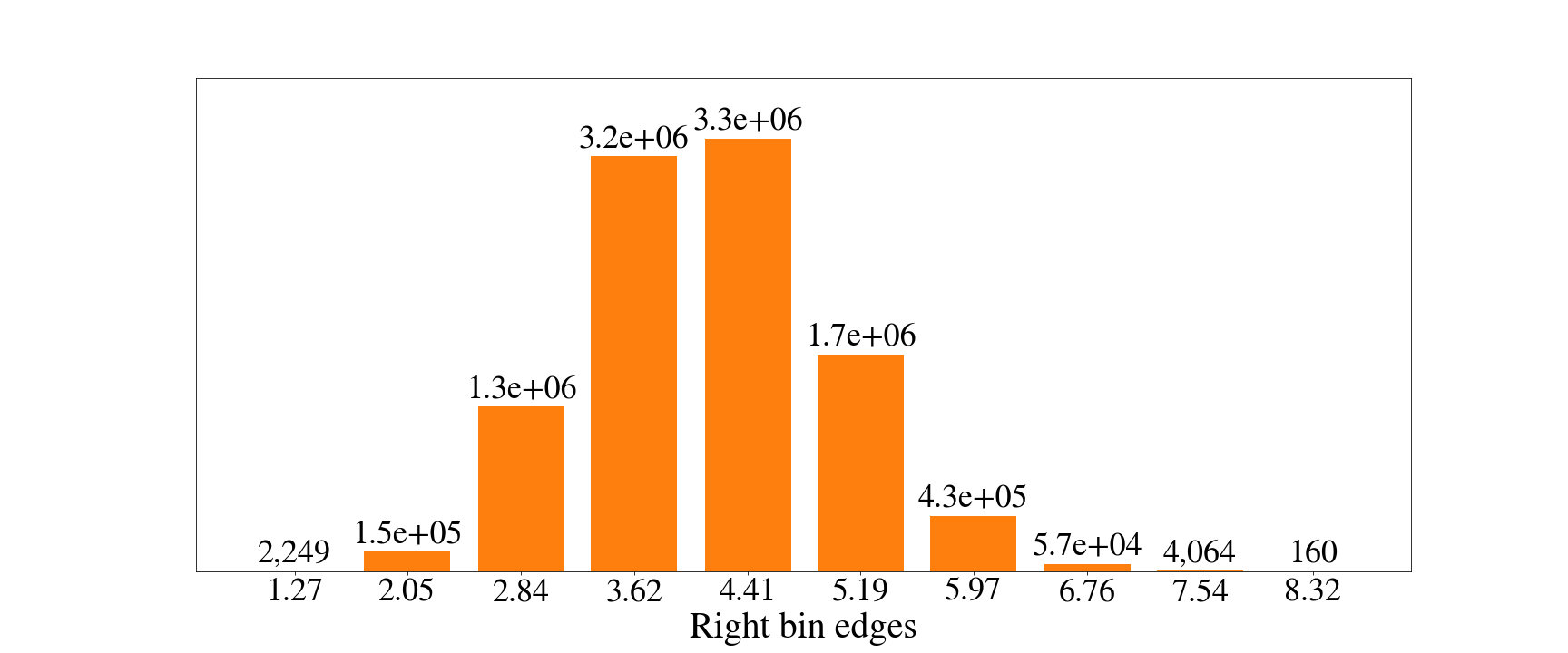}
    \caption{Number of pairwise $z$ distances within each bin for 4,500 random latent embeddings.
    Horizontal axis marks represent the right edge of each bin.
    Data labels represent the count of pairwise distances that fall within each bin.}
    \label{fig:bin_dist_latent}
\end{figure}

In Figures \ref{fig:bin_dist_training} and \ref{fig:bin_dist_latent}, we present the distribution of pairwise $z$ distances for both the training and random latent points.
As mentioned in Section \ref{sec:results}, the final bin in both cases is excluded from the correlation coefficient calculations since it does not contain sufficient observations.

\begin{figure}
    \centering
    \begin{subfigure}{0.6\textwidth}
        \centering
        \includegraphics[width=\linewidth]{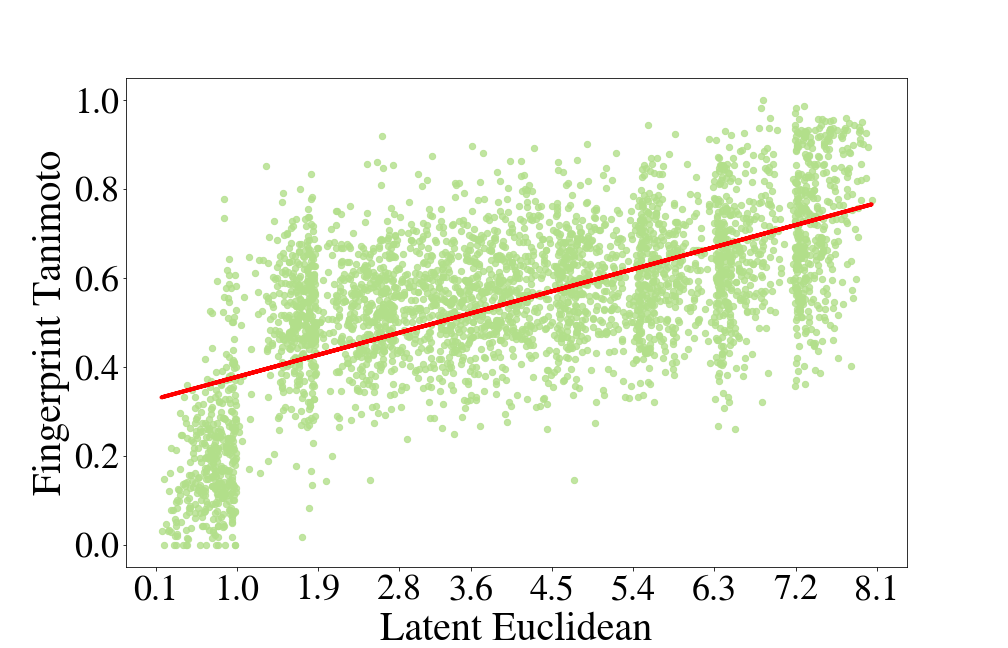}
        \caption{Fingerprint (0.64)}
    \end{subfigure}
    \begin{subfigure}{0.6\textwidth}
        \centering
        \includegraphics[width=\linewidth]{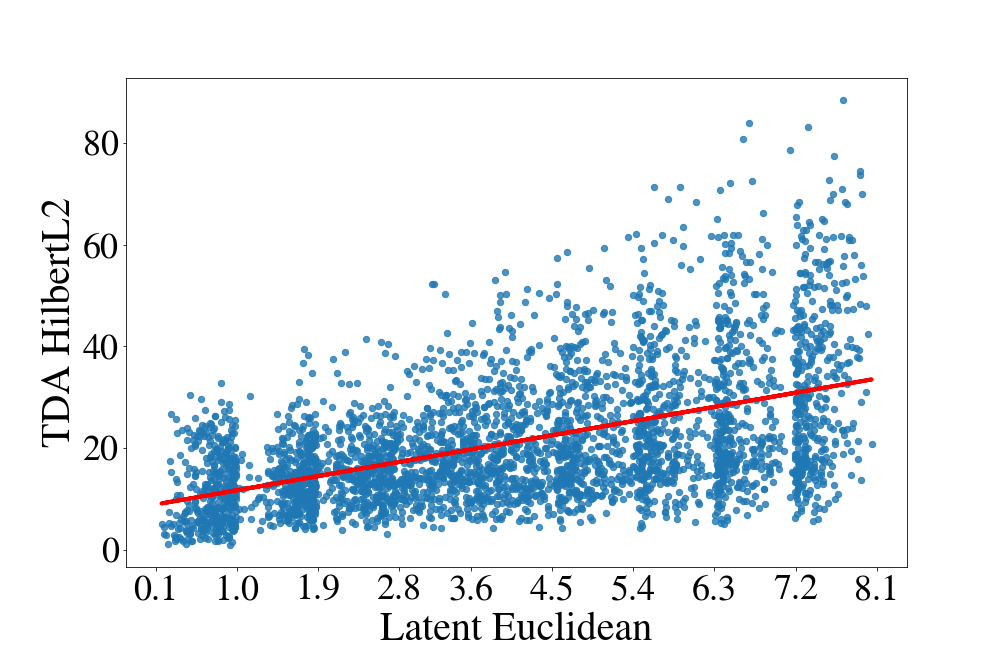}
        \caption{TDA: Partial charge (0.51)}
    \end{subfigure}
    \begin{subfigure}{0.6\textwidth}
        \centering
        \includegraphics[width=\linewidth]{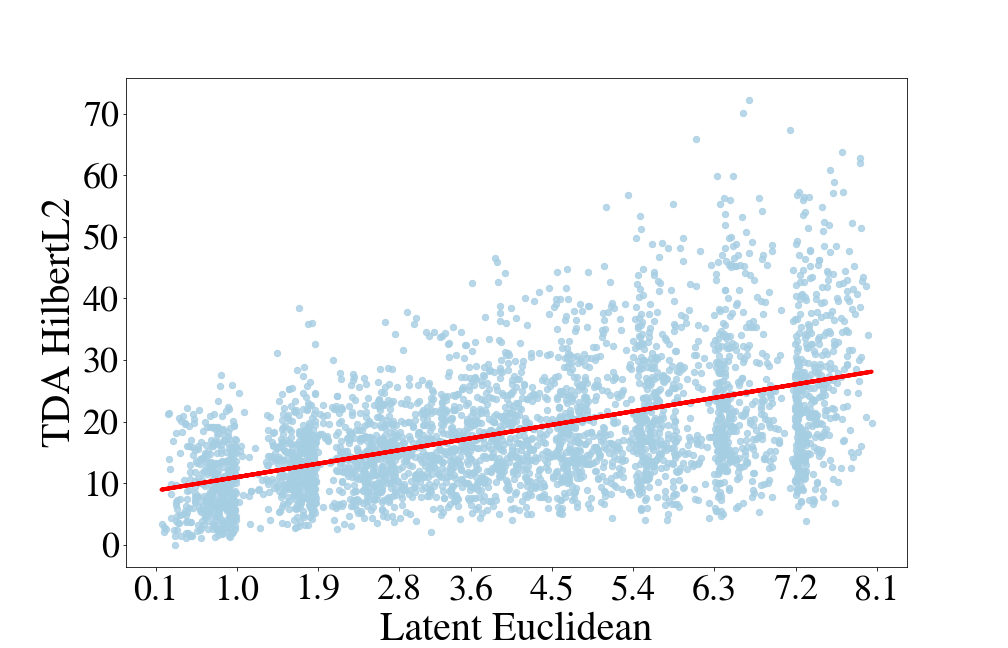}
        \caption{TDA: Radius (0.51)}
    \end{subfigure}
    \caption{Comparison of distances for a single random draw of 3,600 molecule pairs (400 from each of the 9 distance bins considered).
    We compare $z$ distances for \textbf{random training data points} with (a) Tanimoto distances, (b) Hilbert function $\ell_2$ distances for two parameter persistence with partial charge as second parameter, and (c) Hilbert function $\ell_2$ distances for two parameter persistence with atomic radius as second parameter.
    \textit{Pearson r} coefficients for each metric comparison are in parentheses next to each sub-figure title.}
    \label{fig:training}
\end{figure}

\begin{figure}
    \centering
    \begin{subfigure}{0.6\textwidth}
        \centering
        \includegraphics[width=\linewidth]{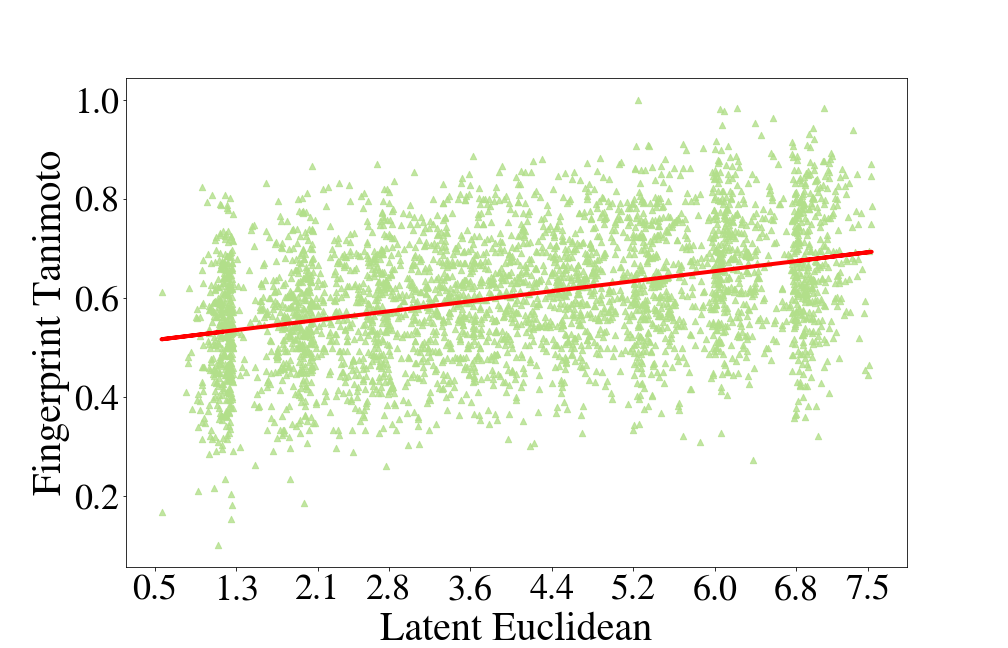}
        \caption{Fingerprint (0.37)}
    \end{subfigure}
    \begin{subfigure}{0.6\textwidth}
        \centering
        \includegraphics[width=\linewidth]{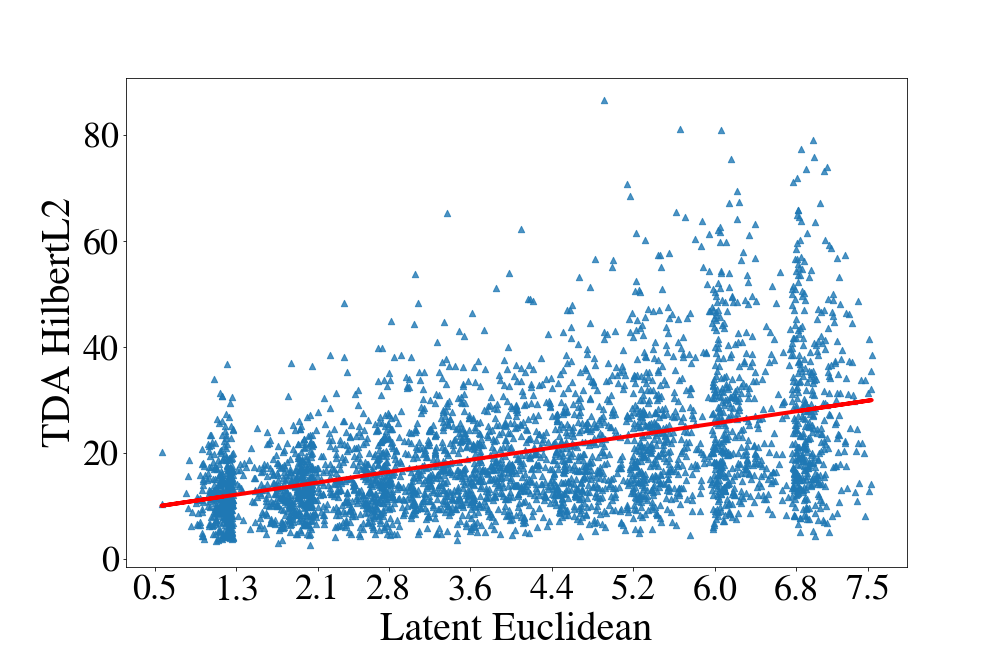}
        \caption{TDA: Partial charge (0.46)}
    \end{subfigure}
    \begin{subfigure}{0.6\textwidth}
        \centering
        \includegraphics[width=\linewidth]{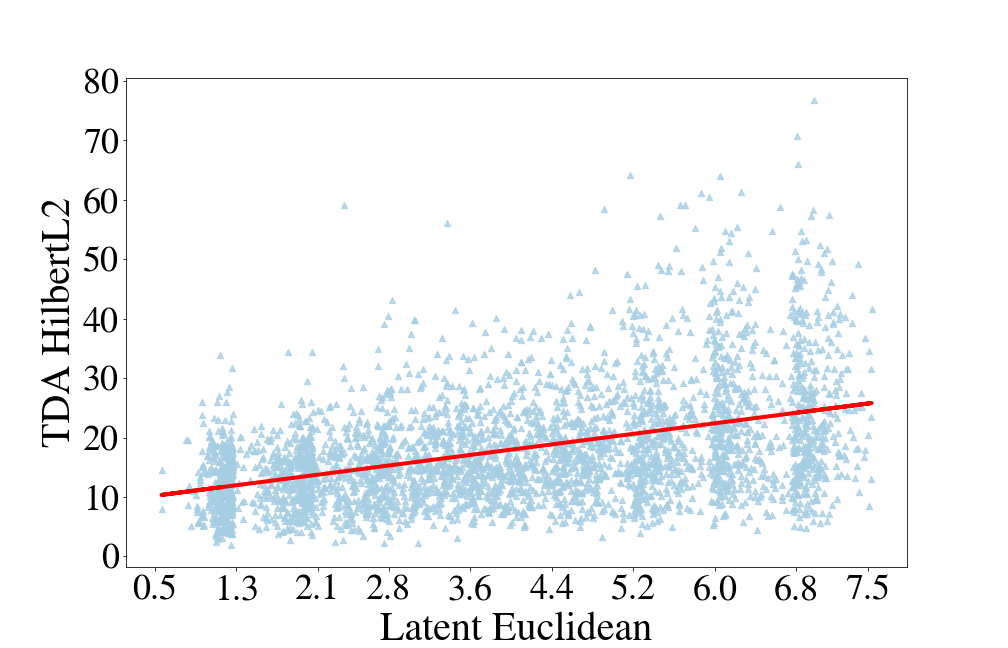}
        \caption{TDA: Radius (0.44)}
    \end{subfigure}
    \caption{Comparison of distances for a single random draw of 3,600 molecule pairs (400 from each of the 9 distance bins considered).
    We compare $z$ distances for \textbf{random samples generated from the latent space} with (a) Tanimoto distances, (b) Hilbert function $\ell_2$ distances for two parameter persistence with partial charge as second parameter, and (c) Hilbert function $\ell_2$ distances for two parameter persistence with atomic radius as second parameter.
    \textit{Pearson r} coefficients for each metric comparison are in parentheses next to each sub-figure title.}
    \label{fig:latent}
\end{figure}

In Figures \ref{fig:training} and \ref{fig:latent}, we plot one random draw, of the 100 displayed in Figure \ref{fig:corr_dist}, and the correlation comparison for $z$ distances with Tanimoto distances and $\ell_2$ Restricted Hilbert function distances. 

\begin{figure}
    \centering
    \begin{subfigure}{0.48\textwidth}
    \centering
    \includegraphics[width=\linewidth]{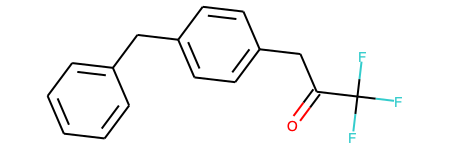}
    \caption{2D molecular drawing for SMILES string \texttt{C1=CC=C(C=C1)CC2=CC=C(C=C2)CC(=O)C(F)(F)F}}
    \end{subfigure}
    \hspace{1em}
    \begin{subfigure}{0.48\textwidth}
    \centering
    \includegraphics[width=\linewidth]{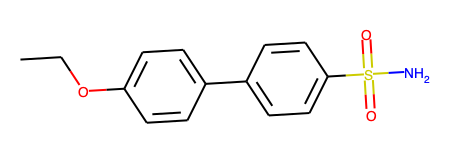}
    \caption{2D molecular drawing for SMILES string \texttt{CCOC1=CC=C(C=C1)C2=CC=C(C=C2)S(=O)(=O)N}}
    \end{subfigure}
    \caption{2D drawing for two molecules that demonstrated higher correlation between latent and persistence diagram metrics than latent and fingerprint metrics.}
    \label{fig:2d_diagram}
\end{figure}

\begin{figure}
    \centering
    \begin{subfigure}{0.7\textwidth}
    \centering
    \includegraphics[width=\linewidth]{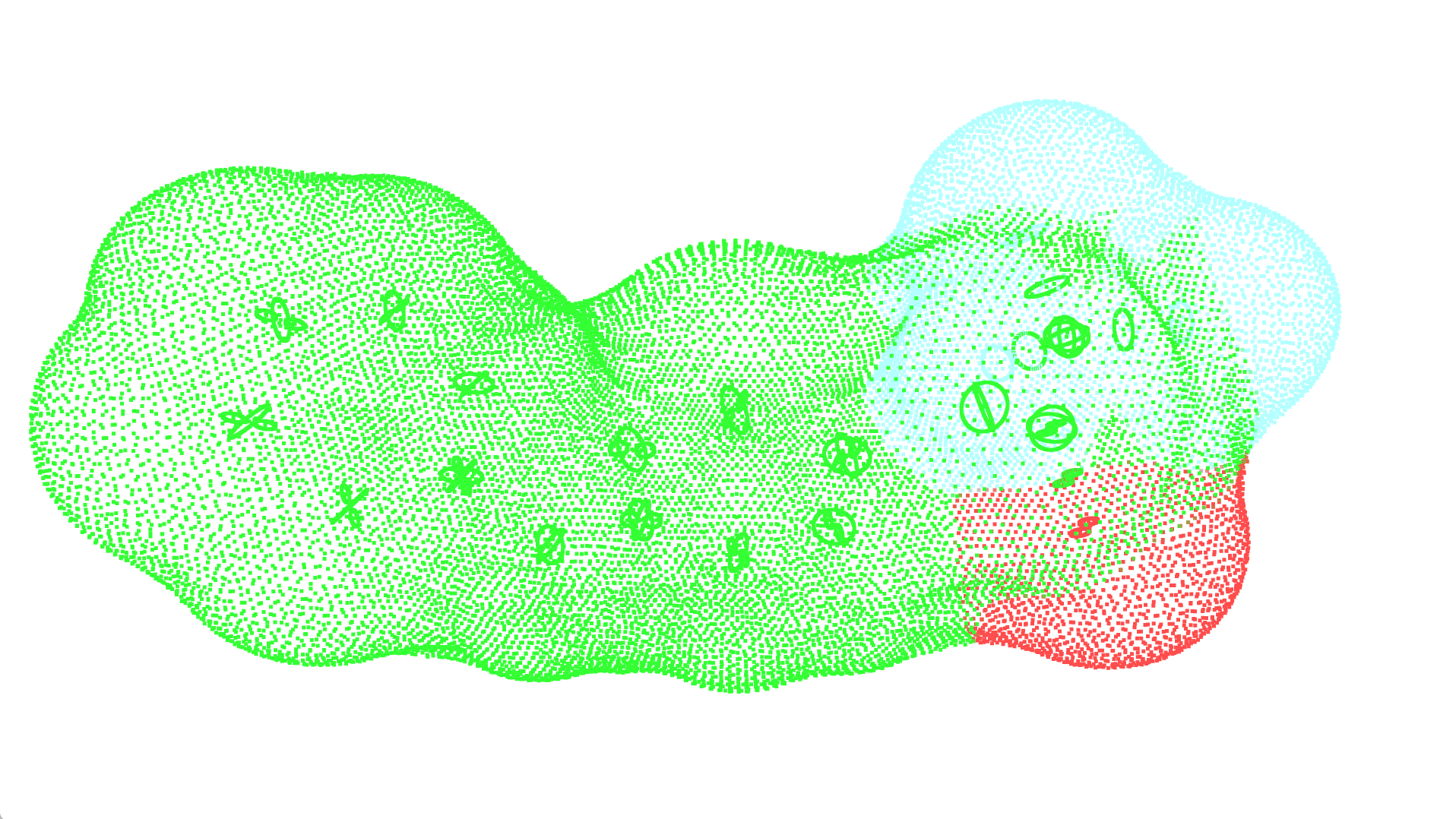}
    \caption{3D point cloud mesh depiction for for SMILES string \texttt{C1=CC=C(C=C1)CC2=CC=C(C=C2)CC(=O)C(F)(F)F}}
    \end{subfigure}
    \begin{subfigure}{0.7\textwidth}
    \centering
    \includegraphics[width=\linewidth]{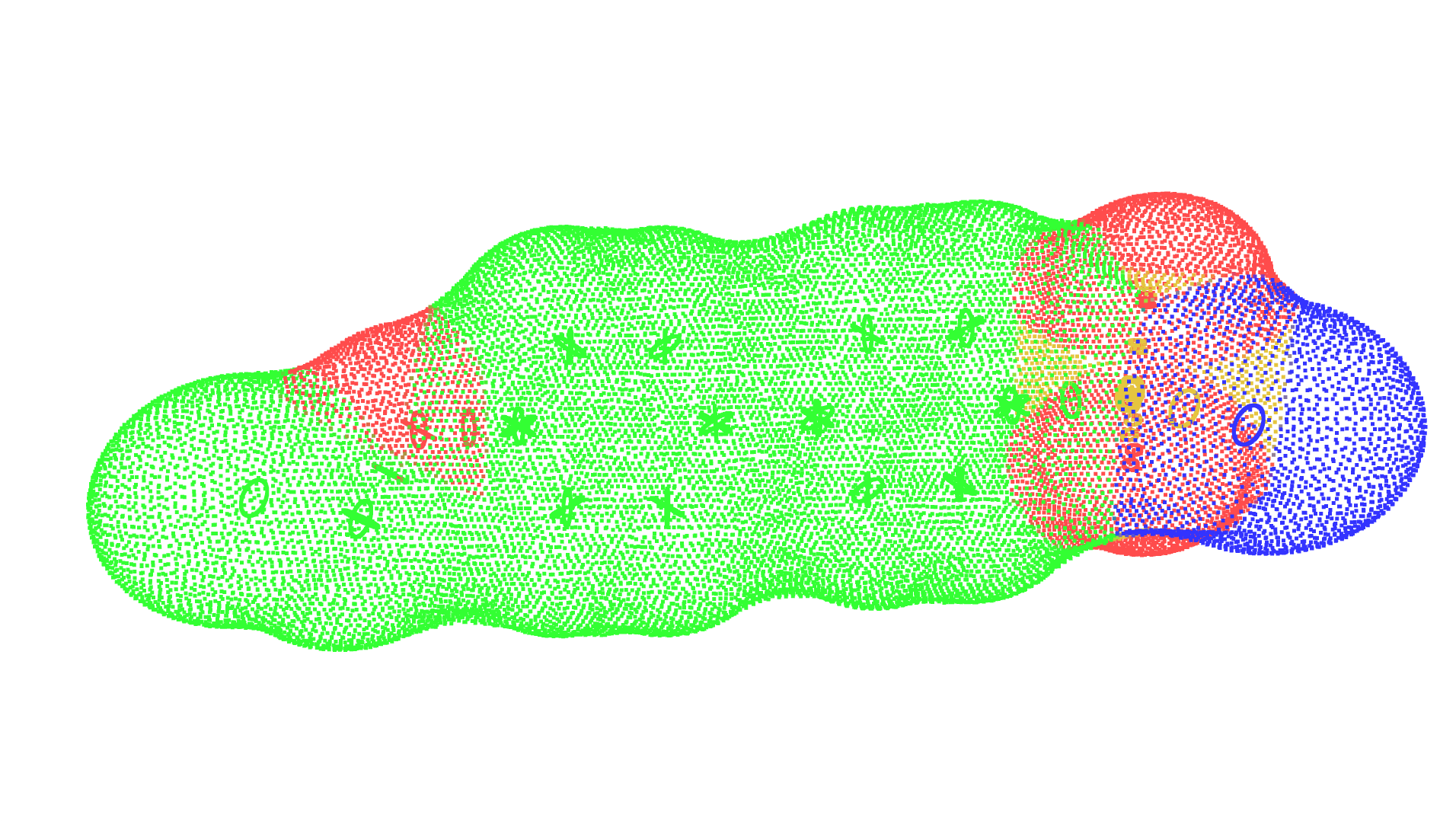}
    \caption{3D point cloud mesh depiction for SMILES string \texttt{CCOC1=CC=C(C=C1)C2=CC=C(C=C2)S(=O)(=O)N}}
    \end{subfigure}
    \caption{3D point cloud mesh depictions for two molecules that demonstrated higher correlation between latent and persistence diagram metrics than latent and fingerprint metrics.}
    \label{fig:3d_diagram}
\end{figure}

\begin{figure}
    \centering
    \begin{subfigure}{0.4\textwidth}
    \centering
    \includegraphics[width=\linewidth]{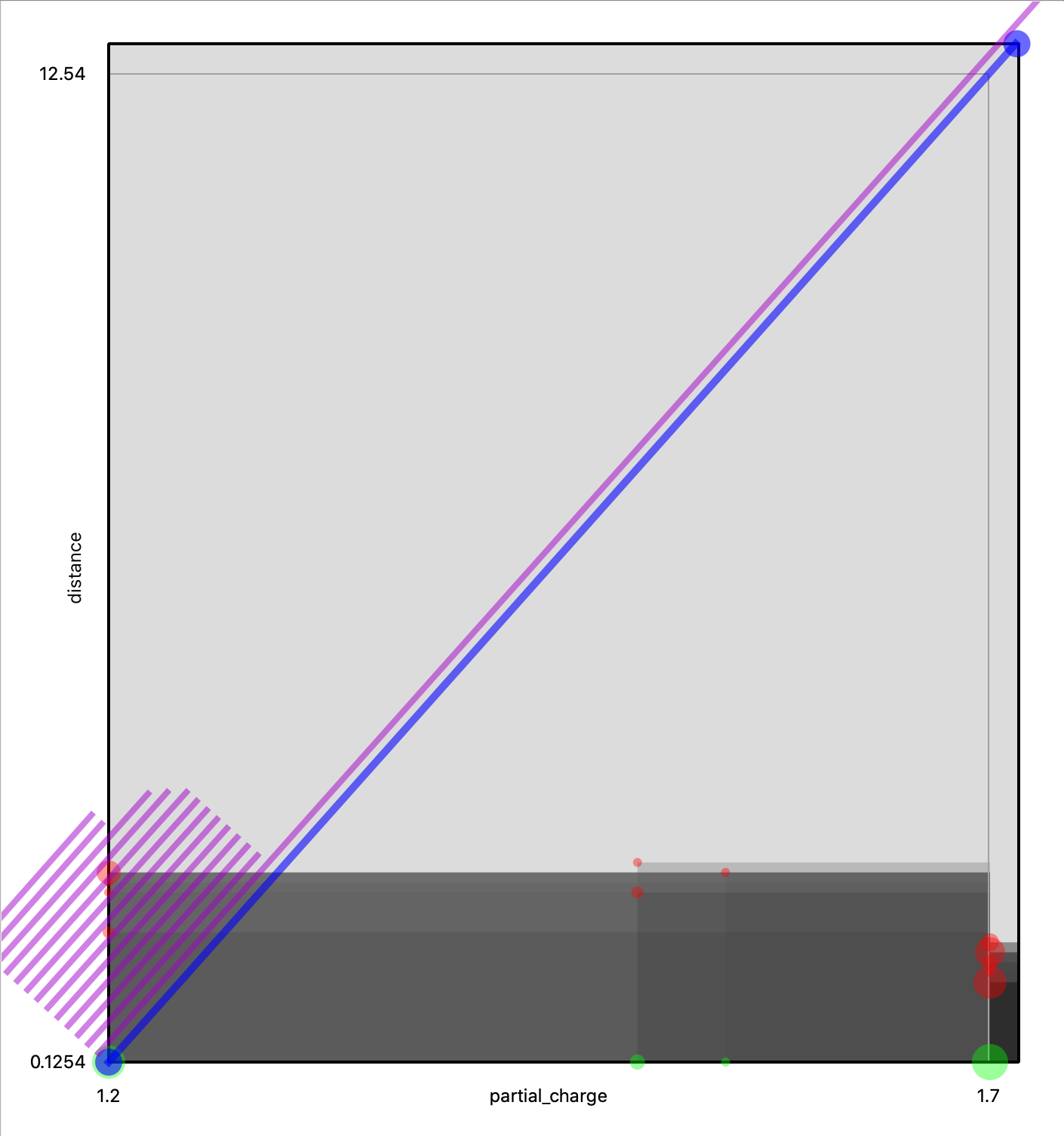}
    \caption{0th Homology two parameter persistence diagram for SMILES string \texttt{C1=CC=C(C=C1)CC2=CC=C(C=C2)CC(=O)C(F)(F)F}.
    Darker shading indicates higher values of the Hilbert function.}
    \end{subfigure}
    \hspace{5em}
    \begin{subfigure}{0.4\textwidth}
    \centering
    \includegraphics[width=\linewidth]{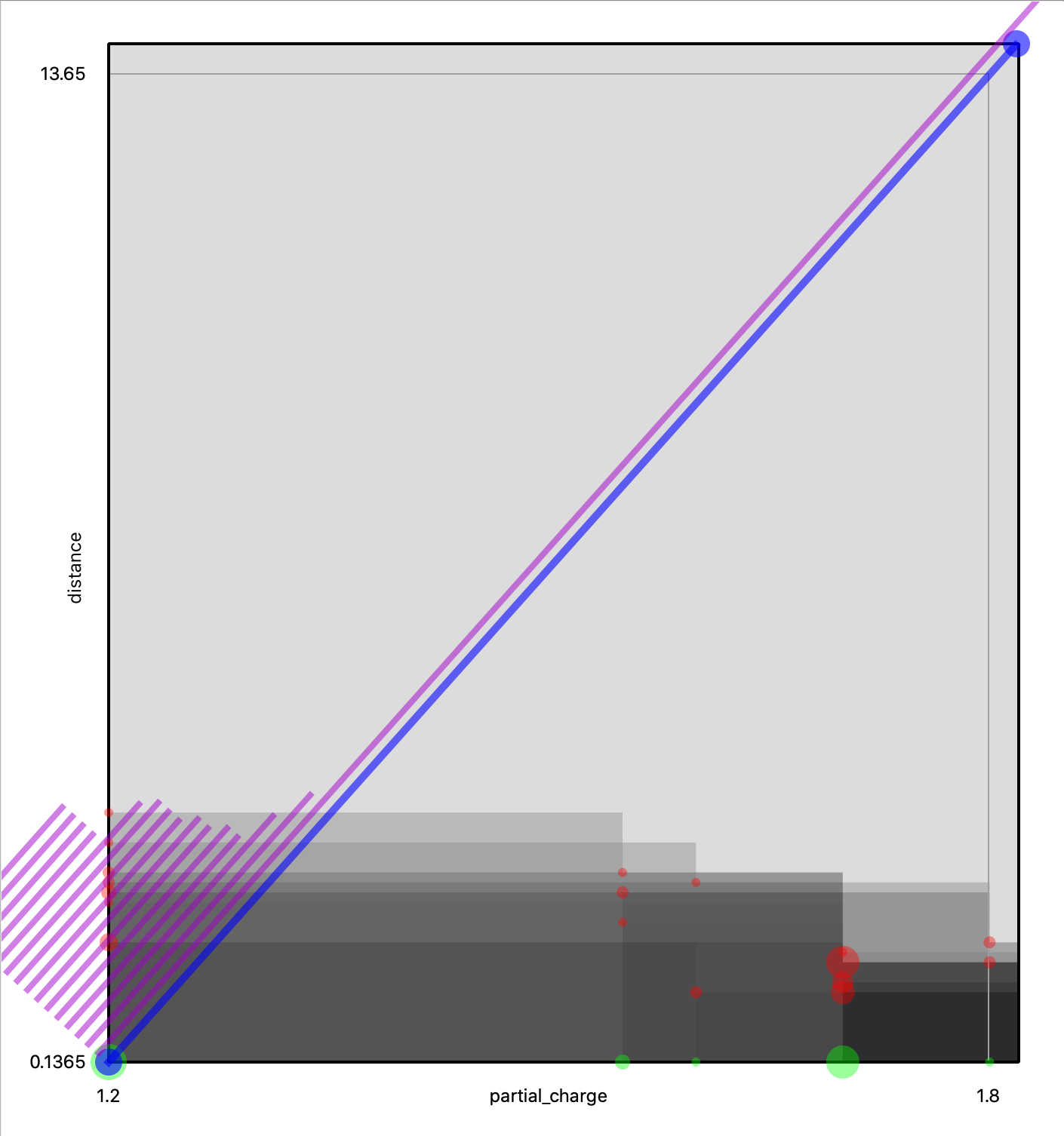}
    \caption{0th Homology two parameter persistence diagram for SMILES string \texttt{CCOC1=CC=C(C=C1)C2=CC=C(C=C2)S(=O)(=O)N}.
    Darker shading indicates higher values of the Hilbert function.}
    \end{subfigure}
    \vspace{5em}
    \begin{subfigure}{0.4\textwidth}
    \centering
    \includegraphics[width=\linewidth]{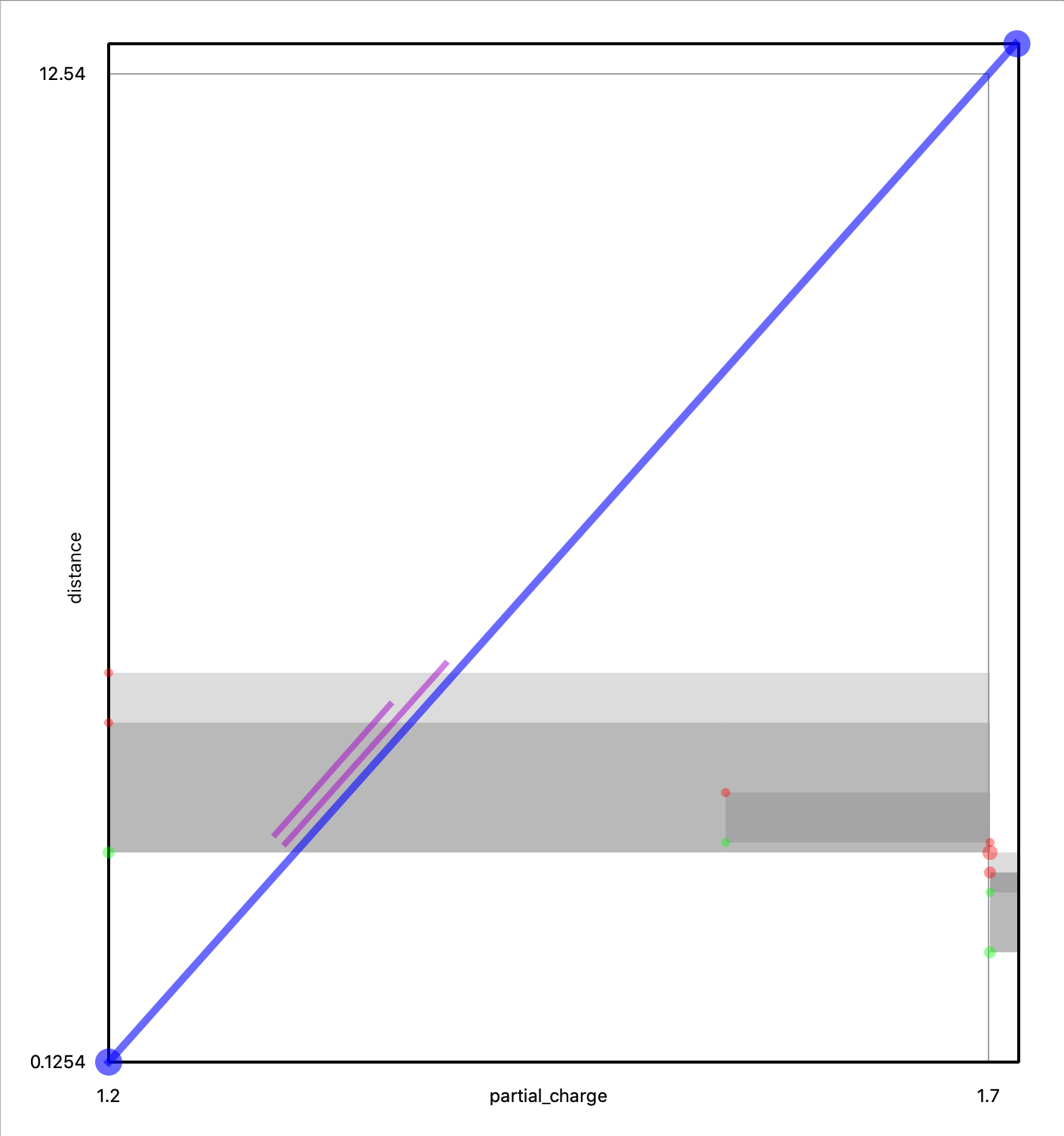}
    \caption{1st Homology two parameter persistence diagram for SMILES string \texttt{C1=CC=C(C=C1)CC2=CC=C(C=C2)CC(=O)C(F)(F)F}.
    Darker shading indicates higher values of the Hilbert function.}
    \end{subfigure}
    \hspace{5em}
    \begin{subfigure}{0.4\textwidth}
    \centering
    \includegraphics[width=\linewidth]{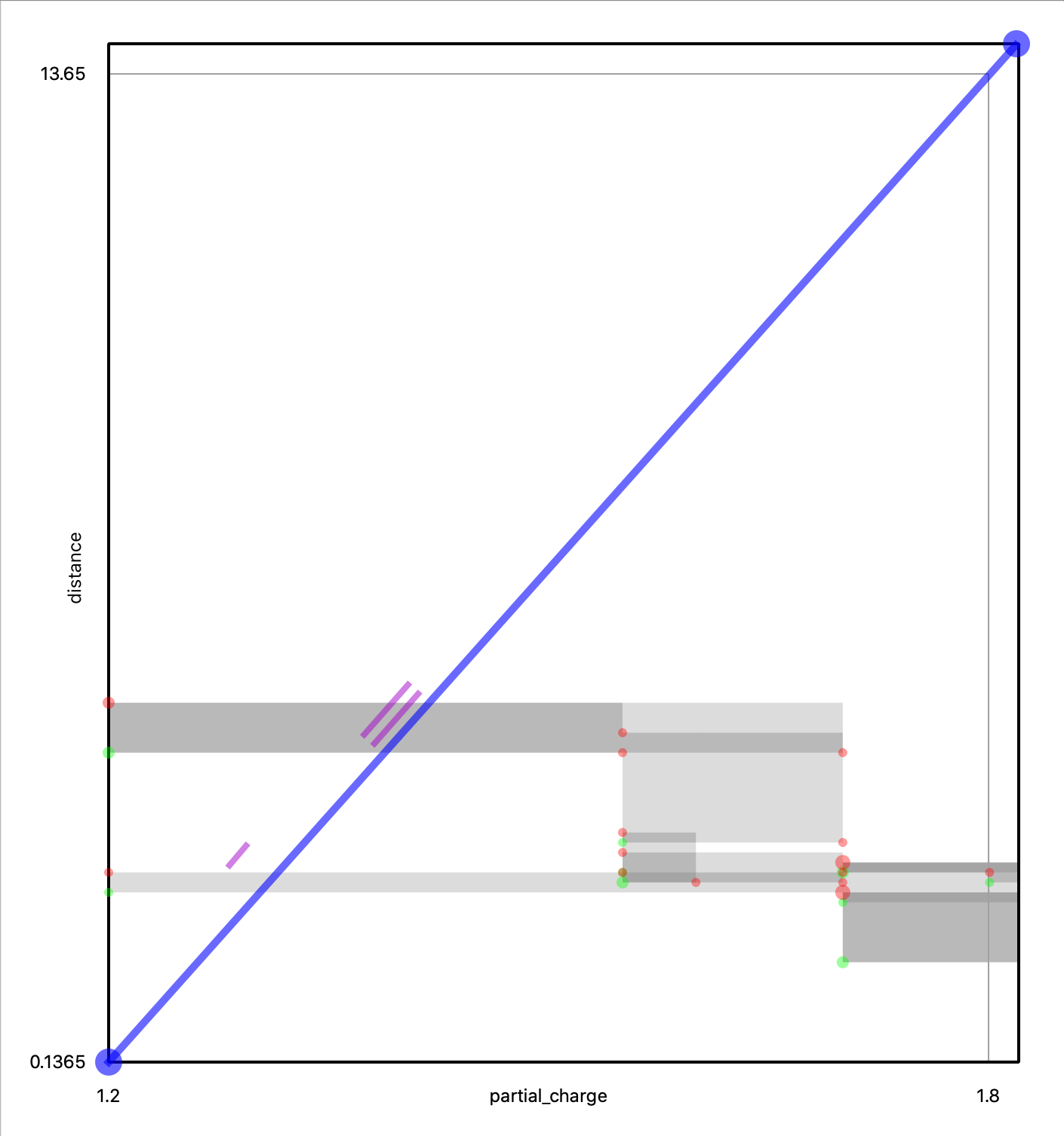}
    \caption{1st Homology two parameter persistence diagram for SMILES string \texttt{CCOC1=CC=C(C=C1)C2=CC=C(C=C2)S(=O)(=O)N}.
    Darker shading indicates higher values of the Hilbert function.}
    \end{subfigure}
    \caption{0th and 1st Homology two parameter persistence diagrams for two molecules that demonstrated higher correlation between latent and persistence diagram metrics than latent and fingerprint metrics.}
    \label{fig:hom_diagram}
\end{figure}

Finally, in Figures \ref{fig:2d_diagram}, \ref{fig:3d_diagram}, and \ref{fig:hom_diagram}, we display two molecules decoded from random samples of the latent distribution that have small $z$ distance (in the 10th percentile of $z$ distance) and similar persistence diagrams (in the 10th percentile of Hilbert function $\ell_2$ distances), but large fingerprint distances (in the 90th percentile of Tanimoto distances).
We choose this molecule pair as a representative example of an instance where proximity in 3D topology representations is well captured in the latent space, but where distance in fingerprint representations do not correlate well with latent distance.
For each molecule, we display a 2D drawing that was created using the RDKit library (Figure \ref{fig:2d_diagram}), a 3D point cloud mesh grid that was created using the MeshLab \cite{LocalChapterEvents:ItalChap:ItalianChapConf2008:129-136}, Open3D \cite{Zhou2018}, and PyMol \cite{PyMOL} libraries (Figure \ref{fig:3d_diagram}), and the two parameter persistence diagram created using the RIVET graphical user interface (Figure \ref{fig:hom_diagram}). 


\end{document}